\documentclass[prx,twocolumn,superscriptaddress,floatfix,longbibliography]{revtex4}
\usepackage{enumerate}
\usepackage{graphicx}
\usepackage{amsmath}
\usepackage{mathtools}
\usepackage{xcolor}
\usepackage{amsfonts}
\usepackage{amssymb}
\usepackage{units}
\usepackage[normalem]{ulem}
\newcommand{\ket}[1]{| #1 \rangle}
\newcommand{\bra}[1]{\langle #1 |}

\begin{document}
\pdfoutput=1
\title{Transmon probe for quantum characteristics of magnons in antiferromagnets}

\author{Vahid Azimi-Mousolou\footnote{Electronic address: v.azimi@sci.ui.ac.ir}}
\affiliation{Department of Applied Mathematics and Computer Science, 
Faculty of Mathematics and Statistics, 
University of Isfahan, Isfahan 81746-73441, Iran}
\affiliation{Department of Physics and Astronomy, Uppsala University, Box 516, 
SE-751 20 Uppsala, Sweden}

\author{Anders Bergman}
\affiliation{Department of Physics and Astronomy, Uppsala University, Box 516, 
SE-751 20 Uppsala, Sweden}

\author{Anna Delin}
\affiliation{Department of Applied Physics, School of Engineering Sciences, 
KTH Royal Institute of Technology, AlbaNova University Center, SE-10691 Stockholm, 
Sweden}
\affiliation{Swedish e-Science Research Center (SeRC), KTH Royal Institute of Technology, 
SE-10044 Stockholm, Sweden}

\author{Olle Eriksson}
\affiliation{Department of Physics and Astronomy, Uppsala University, Box 516, 
SE-751 20 Uppsala, Sweden}

\author{Manuel Pereiro}
\affiliation{Department of Physics and Astronomy, Uppsala University, Box 516, 
SE-751 20 Uppsala, Sweden}

\author{Danny Thonig}
\affiliation{School of Science and Technology, \"Orebro University, SE-701 82, 
\"Orebro, Sweden}

\author{Erik Sj\"oqvist\footnote{Electronic address: 
erik.sjoqvist@physics.uu.se}}
\affiliation{Department of Physics and Astronomy, Uppsala University, 
Box 516, SE-751 20 Uppsala, Sweden}

\date{\today}% It is always \today, today,

\begin{abstract}
The detection of magnons and their quantum properties, especially in antiferromagnetic (AFM) materials, is a substantial step to realize many ambitious advances in the study of nanomagnetism and the development of energy efficient quantum technologies.
The recent development of hybrid systems based on superconducting circuits provides the possibility of engineering quantum sensors that exploit different degrees of freedom. Here, we examine the magnon-photon-transmon hybridisation based on bipartite AFM  
materials, which gives rise to an effective coupling between a transmon qubit and magnons in a bipartite AFM. We demonstrate how magnetically invisible magnon modes, their chiralities and quantum properties such as nonlocality and two-mode magnon entanglement in bipartite AFMs can be characterized through the Rabi frequency of the superconducting transmon qubit. 
\end{abstract}
\maketitle
\section{Introduction}
During the last decade, there have been considerable advancements in the use of  magnons for storing, transmitting, and processing information. 
This rapid progress has turned the emerging research field of magnonics into a promising candidate for innovating information processing technologies \cite{barman2021}. 
The combination of magnonics with quantum information processing provides a highly interdisciplinary physical platform for studying various quantum phenomena in 
spintronics, quantum electrodynamics, and quantum information science. Indeed, the quantum magnonics exhibits distinct quantum properties, which can be utilized for 
multi-purpose quantum tasks \cite{awschalom2021, yli2020, lachance-quirion2019, clerk2020, yuan2022}.

Despite significant progress in quantum magnonics \cite{awschalom2021, yli2020, lachance-quirion2019, clerk2020, yuan2022, Lachance-Quirion2020, azimi-mousolou2020, azimi-mousolou2021, liu2022,li2018a, li2019, zhang2019, bossini2019, yuan2020a, yuan2020b, tabuchi2014, yuan2017, xiao2019, johansen2018}, there are still many features and challenges that need to be addressed in theory and in the laboratory. In particular, the experimental 
verification of non-classical magnon states and quantum properties such as squeezed and entangled states would pave the way
for many possible research strategies.
The key point is interconnections between magnetic materials and electronic quantum systems. Superconducting qubits have been successfully used to detect magnons in ferromagnetic materials \cite{Lachance-Quirion2020}.
However, antiferromagnetic (AFM) materials are more sustainable for quantum applications as they offer lower magnetic susceptibility, faster dynamics, smaller device features and lower energy consumption compared to ferromagnetic materials \cite{barman2021}.
Recently, we have theoretically examined magnon-magnon entanglement and squeezing in AFMs \cite{azimi-mousolou2020, azimi-mousolou2021, liu2022}.  

Here, we examine the possibility to 
combine the advantageous features of transmon and AFM materials. To this end, we demonstrate effective coupling between a superconducting transmon qubit and a bipartite AFM material. We show how the polarized (chiral)  magnons and bipartite magnon-magnon entanglement in the AFM can be detected through the measurement of Rabi frequency of the transmon qubit. The proposed setup is suitable for the experimental study of the quantum properties of magnons in a wide range of crystalline and synthetic AFM materials, such as NiO and MnO, MnF$_2$ and FeF$_2$, two-dimensional Ising systems like MnPSe$_3$, YIG-based synthetic AFMs, and perovskite manganites \cite{Jie2018,Takashi2016,Haakon2019,Thuc2021,Sheng2021, Changting2021, Rini2007, Ulbrich2011, rezende2019}.

The outline of the paper is as follows : In sec. \ref{Magnon-Photon-Transmon-hybridization} we describe  magnon-photon-transmon hybridization and derive the interacting Hamiltonian. In sec. \ref{magnon-magnon-entanglement}, we discuss two-mode magnon entanglement in AFM materials. In sec. \ref{Sensing-magnons-quantum-characteristics-with-transmons}, we obtain an effective magnon-transmon coupling and show how this effective coupling mechanism allows to experimentally study quantun charachteristics  of magnons in antiferromagnetic materials. The paper ends with a conclusion in sec. \ref{conclusion}. 

\section{Magnon-Photon-Transmon hybridization}
\label{Magnon-Photon-Transmon-hybridization}
\begin{figure}[h]
\begin{center}
\includegraphics[width=80mm]{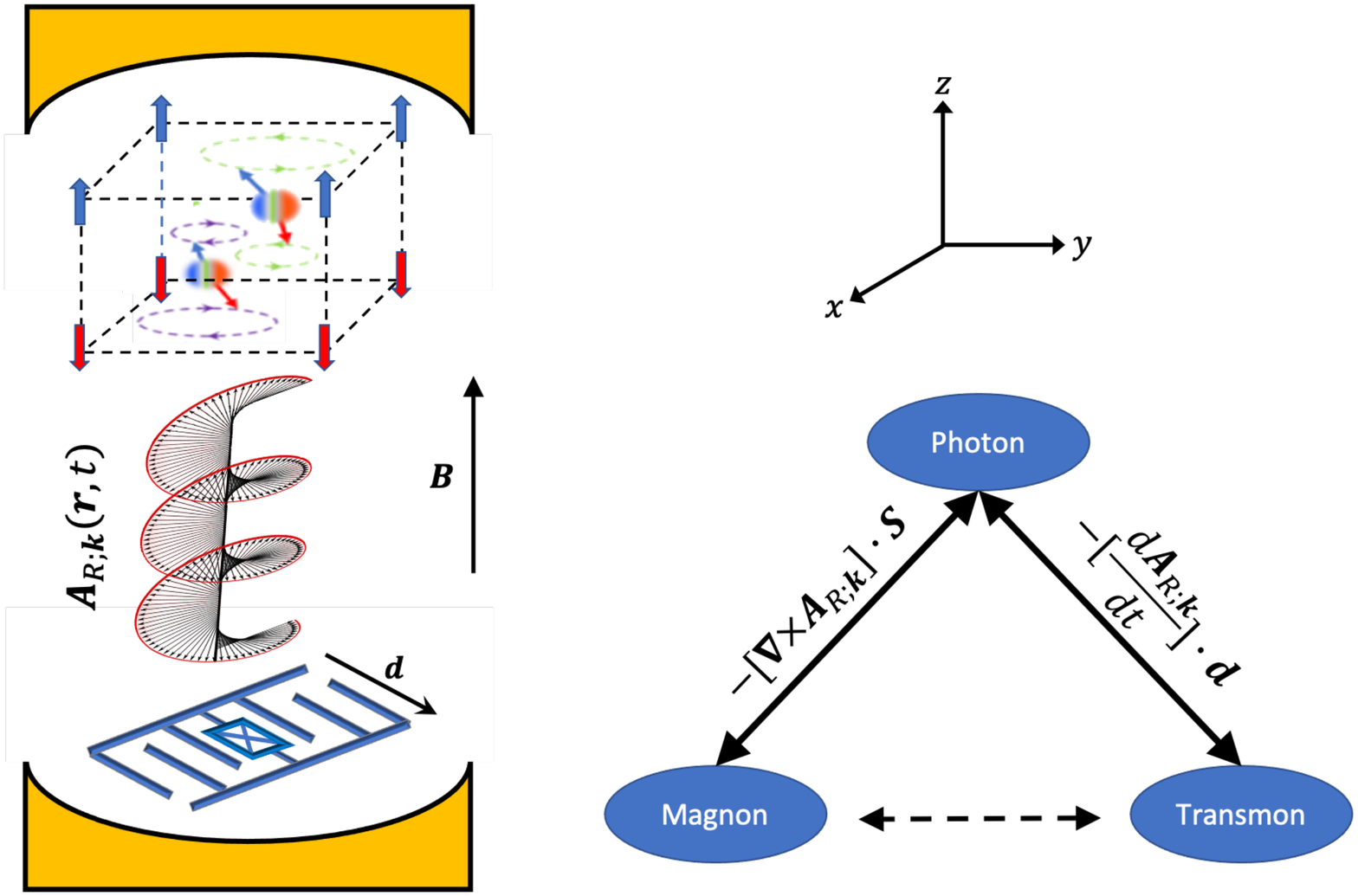}
\end{center}
\caption{(Color online) Schematic illustration of magnon-photon-transmon hybridization. A circularly polarized microwave cavity electromagnetic field, which is described by the vector potential $\mathbf{A}_{R; \mathbf{k}}(\mathbf{r}, t)$, can interact with magnons in an antiferromagnetic material and a superconducting transmon qubit. The cavity walls are illustrated with yellow segments in the left panel. An antiferromagnetic material hosts two chiral magnons, which are shown with three-color balls in the cubic lattice inside the cavity. Two magnos are degenerate in the absence of magnetic field and a small external magnetic field $\mathbf{B}$ in the $z$ direction breaks this degenerecy (see also Fig. \ref{fig:energy}). While the coupling between magnon and cavity filed is achieved through magnetic-dipole interaction, an electric-dipole interaction describe coupling between cavity filed and transmon (right panel).}
\label{fig:model}
\end{figure}
\begin{figure}[h]
\begin{center}
\includegraphics[width=90mm]{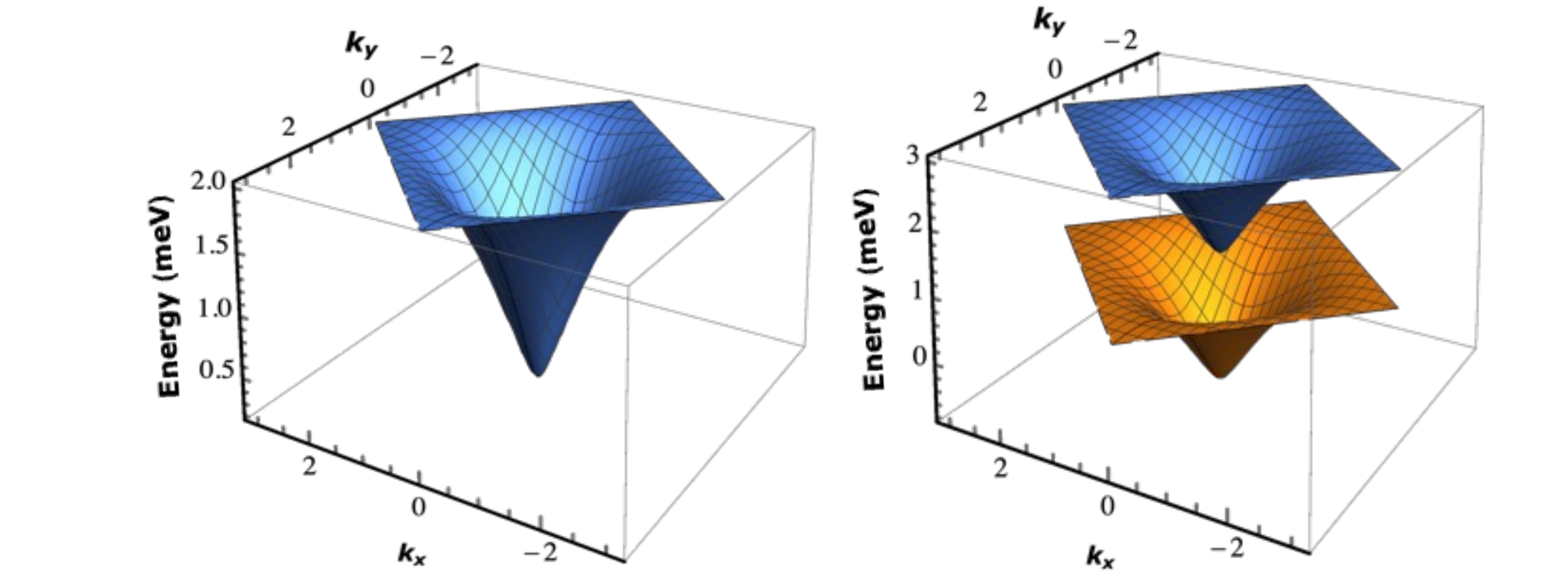}
\end{center}
\caption{(Color online) Magnon energy dispersions $\omega_{\alpha_{\mathbf{k}}}$ and $\omega_{\beta_{-\mathbf{k}}}$ in the first Brillion zone of a square lattice with lattice constant $a=1$ for an easy-axis AFM. 
As model parameters, we use $|J|=1 meV$ for antiferromagnetic Heisenberg exchange, $\mathcal{K}_z=0.01J$ for uniaxial anisotropy, and $S=1/2$.
Two magnons are degenerate in the absence of an external magnetic field $\mu_B B=0$ (left panel). A magnetic field $\mu_B B=1 meV$ in the $z$ direction breaks the degeneracy (right panel).}
\label{fig:energy}
\end{figure}

In this section, we describe a photon-mediated coupling mechanism between a superconducting transmon qubit and polarized  magnons in a bipartite AFM. 
We assume a hybrid system composed of a single crystal or synthetic AFM, a transmon-type superconducting qubit, and a microwave cavity, as illustrated in Fig.~\ref{fig:model}. The system hosts four modes including two magnon modes in an AFM compound, a transmon qubit, and a microwave cavity electromagnetic mode. The dynamics of the hybridized magnon-photon-transmon system can be described by the Hamiltonian
\begin{eqnarray}
H=H_{\text{m}}+H_{\text{ph}}+H_{\text{m-ph}}+H_{\text{q}}+H_{\text{ph-q}},
\label{MH}
\end{eqnarray}
where the term $H_{\text{m}}$ describes the magnon subsystem, 
$H_{\text{ph}}$ describes the microwave photon, 
$H_{\text{m-ph}}$ describes the magnon-photon interaction, 
$H_{\text{q}}$ describes the transmon and 
$H_{\text{ph-q}}$ describes the photon-transmon interaction. 
They are described in detail as follows:

{\it Two-mode magnon system}: $H_{\text{m}}$ represents a two-mode magnon Hamiltonian in a bipartite treatment of an AFM material. 
Consider an AFM spin Hamiltonian $\sum_{i, j}\mathbf{S}_{i}\mathbb{I}_{ij}\mathbf{S}_{j}+\sum_{i}\mathbf{B}\cdot\mathbf{S}_{i}$, where $\mathbf{S}_{i}$ is the spin vector operator at lattice site $i$, $\mathbb{I}_{ij}$ is the bi-linear interaction tensor matrix between sites $i$ and $j$, and $\mathbf{B}$ is an external field. By applying the Holstein-Primakoff transformation at low temperature followed by the Fourier transformation to the AFM spin Hamiltonian, $H_{\text{m}}$ can be described in terms of a pair of interacting collective bosonic modes in the lattice momentum $\mathbf{k}$-space as \cite{azimi-mousolou2020, azimi-mousolou2021} (we assume $\hbar=1$ throughout the paper)

\begin{eqnarray}
H_{\text{m}}^{\mathbf{k}} &=& 
\omega_{a_{\mathbf{k}}}a_{\mathbf{k}}^{\dagger} a_{\mathbf{k}} + 
\omega_{b_{-\mathbf{k}}}b_{-\mathbf{k}}^{\dagger} b_{-\mathbf{k}}\nonumber\\
&& + g_{\text{m-m}}^{\mathbf{k}} a_{\mathbf{k}} b_{-\mathbf{k}} + \left(g_{\text{m-m}}^{\mathbf{k}}\right)^{*}a_{\mathbf{k}}^{\dagger}b_{-\mathbf{k}}^{\dagger}.
\label{MMH}
\end{eqnarray}
The $a_{\mathbf{k}}^{\dagger}$ ($a_{\mathbf{k}}$) and $b^{\dagger}_{-\mathbf{k}}$ ($b_{-\mathbf{k}}$) are bosonic creation (annihilation) operators on the two sublattices $A$ and $B$ with opposite magnetizations in the bipartite AFM.
Bosonic operators on opposite sublattices commute and define a pair of interacting magnons in the Kittel $(a, b)$ modes.
The Kittel modes can be hybridized into the diagonal magnon modes $(\alpha, \beta)$ through the SU(1,1) Bogoliubov transformation
 \begin{eqnarray}
\left(
\begin{array}{cc}
  a_{\mathbf{k}}    \\
   b_{-\mathbf{k}}^{\dagger}       
\end{array}
\right)=\left(
\begin{array}{cc}
  u_{\mathbf{k}}& v_{\mathbf{k}}    \\
v_{\mathbf{k}}^{*}& u_{\mathbf{k}}^{*}       
\end{array}
\right)\left(
\begin{array}{cc}
  \alpha_{\mathbf{k}}    \\
   \beta_{-\mathbf{k}}^{\dagger}       
\end{array}
\right),
\label{eq:FBT}
\end{eqnarray}
where $u_{\mathbf{k}} =\cosh(r_{\mathbf{k}})$ and $v_{\mathbf{k}} = \sinh(r_{\mathbf{k}})e^{i\phi_{\mathbf{k}}}$ with 
\begin{eqnarray}
r_{\mathbf{k}}&=&\tanh^{-1}\left[\frac{1-\sqrt{1-|\Gamma_{\mathbf{k}}|^{2}}}{|\Gamma_{\mathbf{k}}|}\right]\ge 0,\nonumber\\
\phi_{\mathbf{k}}&=&\pi-\arg[\Gamma_{\mathbf{k}}],\ \ \ \ 
\Gamma_{\mathbf{k}}=\frac{2g_{\text{m-m}}^{\mathbf{k}}}{\omega_{a_{\mathbf{k}}}+\omega_{b_{-\mathbf{k}}}}.
\label{r-phi}
\end{eqnarray}
In terms of the $(\alpha, \beta)$ modes, the magnon Hamiltonian $H_{\text{m}}^{\mathbf{k}}$ takes the diagonal form
 \begin{eqnarray}
H_{\text{m}}^{\mathbf{k}} = 
\omega_{\alpha_{\mathbf{k}}}\alpha_{\mathbf{k}}^{\dagger}\alpha_{\mathbf{k}} +
\omega_{\beta_{-\mathbf{k}}}\beta_{-\mathbf{k}}^{\dagger} \beta_{-\mathbf{k}}.
\label{DMMH}
\end{eqnarray}
The bosonic diagonal modes $\alpha$ and $\beta$ describe two right and left circularly polarized (chiral) magnons \cite{barman2021, zhang2020}, which are degenerate in the absence of an external magnetic field. As shown in Fig.\ \ref{fig:energy}, for a system with only diagonal components of $\mathbf{J}_{ij}$ (=J), a magnetic field in the $z$ direction, i.e., parallel to the magnetization of the two sublattices, breaks the degeneracy \cite{azimi-mousolou2020, azimi-mousolou2021}.

{\it Microwave photon}: For the second term of the hybrid Hamiltonian in Eq.~\eqref{MH}, we assume a right circularly polarized microwave cavity electromagnetic field with the single cavity mode frequency $\omega_{c_{\mathbf{k}}}$ \cite{yuan2017, xiao2019, zhang2020, azimi-mousolou2021}. This is described by the vector potential 
\begin{eqnarray}
\mathbf{A}_{R; \mathbf{k}}(\mathbf{r}, t)&=&A_{0}\left[\mathbf{e}_{R}c_{\mathbf{k}}e^{-i(\mathbf{k} \cdot \mathbf{r}+\omega_{c_{\mathbf{k}}} t)}+\mathbf{e}^{*}_{R}c^{\dagger}_{\mathbf{k}}e^{i(\mathbf{k} \cdot \mathbf{r}+\omega_{c_{\mathbf{k}}} t)}\right]
\nonumber\\
&=&e^{it\omega_{c_{\mathbf{k}}} c^{\dagger}_{\mathbf{k}}c_{\mathbf{k}}}\mathbf{A}_{R; \mathbf{k}}(\mathbf{r}, 0)e^{-it\omega_{c_{\mathbf{k}}} c^{\dagger}_{\mathbf{k}}c_{\mathbf{k}}}.
\end{eqnarray}
The vector $\mathbf{k}$ is the propagation direction of the electromagnetic wave,  $A_{0}$ is the amplitude of the vector potential, and $c_{\mathbf{k}} (c^{\dagger}_{\mathbf{k}})$ is the annihilation (creation) operator of the right circularly polarized photon with unit vector $\mathbf{e}_{R}=\frac{1}{\sqrt{2}}(1, -i, 0)$. Both $\omega_{c_{\mathbf{k}}}$ and $A_{0}$ can be tuned by changing the volume of the cavity and the separation 
distance between the two conductor plates in the cavity. Here, we focus on the lowest energy cavity mode and disregard contributions from the higher energy cavity modes. In the rotating frame, the photon contribution to the full Hamiltonian in Eq.~\eqref{MH} is
\begin{eqnarray}
H_{\text{ph}}^{\mathbf{k}} =\omega_{c_{\mathbf{k}}}c^{\dagger}_{\mathbf{k}}c_{\mathbf{k}},
\label{PhH}
\end{eqnarray}
for a given $\mathbf{k}$.

{\it Magnon-Photon interaction}: By turning on the electromagnetic field, the magnon modes start to interact with the cavity mode through the magnetic-dipole coupling. Explicitly, the electromagnetic field induces a magnetic field $\mathbf{B}_{\text{ph}}$, 
which interacts with the total spin $\mathbf{S}$ of the AFM material through the Zeeman interaction term \cite{yuan2017, xiao2019, zhang2020, azimi-mousolou2021}
\begin{eqnarray}
H_{\text{m-ph}}=-\mathbf{B}_{\text{ph}}\cdot\mathbf{S}.
\end{eqnarray}
In the rotating frame, the photon-induced magnetic field is given by $\mathbf{B}_{\text{ph}}=\nabla\times \mathbf{A}_{\mathbf{k}}(\mathbf{r}, 0)$. Following the bosonization procedure used to derive the Hamiltonian $H_{\text{m}}^{\mathbf{k}}$, we obtain  
the bosonized resonant magnon-photon interaction Hamiltonian  
\begin{eqnarray}
H_{\text{m-ph}}^{\mathbf{k}}=-g_{\text{m-ph}}^{\mathbf{k}}c^{\dagger}_{\mathbf{k}}\alpha_{\mathbf{k}} + {\rm H.c.}
\label{BCH}
\end{eqnarray}
%for the vector $\mathbf{k}$ fixed along the $(0, 0, 1)$ direction. 
The  
off-resonant interaction $(-g_{\text{m-ph}}^{\mathbf{k}}c_{\mathbf{k}}\beta_{-\mathbf{k}}+ {\rm H.c.})$ is neglected due to energy conservation.
Here, the magnon-photon exchange coupling is 
\begin{eqnarray}
g_{\text{m-ph}}^{\mathbf{k}}=\lambda_{\mathbf{k}}(u_{\mathbf{k}}+v^{*}_{\mathbf{k}})
\end{eqnarray}
with $\lambda_{\mathbf{k}}=A_{0}k\sqrt{S}$ and we choose to study the case when $\mathbf{k}=(0, 0, k)$.

{\it Transmon qubit}: The third subsystem consists of a superconducting qubit described by the Hamiltonian \cite{koch2007} 
\begin{eqnarray}
H_{\text{q}}=4E_{C}\hat{n}^{n}-E_{J}\cos\hat{\phi},
\label{TQhamiltonian}
\end{eqnarray}
where the first term corresponds to the kinetic energy contribution from a capacitor and the second term is the potential energy contribution by a Josephson junction. At a sufficiently large  $E_{J}/E_{C}$, the superconducting system enters the transmon qubit  regime. Following the ladder operator approach, one may represent the momentum, $\hat{n}$, and position, $\hat{\phi}$, operators in terms of bosonic annihilation (creation) operator $\eta$ ($\eta^{\dagger}$) as 

\begin{eqnarray}
\hat{n}&=&i\left(\frac{E_{J}}{32E_{C}}\right)^{1/4}(\eta^{\dagger}-\eta),\nonumber\\
\hat{\phi}&=&\left(\frac{2E_{C}}{E_{J}}\right)^{1/4}(\eta^{\dagger}+\eta).
\end{eqnarray}
By using the ladder representation, one can write the Hamiltonian in Eq.\ \eqref{TQhamiltonian} in the form of the following anharmonic oscillator Hamiltonian 
\begin{eqnarray}
H_{\text{q}}\approx\left[\omega_q+\frac{\xi}{2}\right]\eta^{\dagger}\eta-\frac{\xi}{2}(\eta^{\dagger}\eta)^{2}.
\end{eqnarray}
This follows from a Taylor expansion of the potential energy term in Eq.\ \eqref{TQhamiltonian} and a rotating wave approximation. 
Here, $\omega_q=\sqrt{8E_{C}E_{J}}-E_{C}$ defines the Rabi transition frequency between the ground state $\ket{g}$ and the first excited state $\ket{e}$, $\xi=E_{C}$ is the anharmonicity. In the transmon regime, the anharmonicity is negative and large enough 
that allows one to focus on the two lowest energy levels of the anharmonic oscillator as a transmon qubit, the Hamiltonian of which can be conveniently reduced to  
\begin{eqnarray}
H_{\text{q}}=\omega_{q}\eta^{\dagger}\eta.
\end{eqnarray}

{\it Photon-transmon interaction}: The large electric dipole of the superconducting qubit, $\hat{\mathbf{d}}=\mathbf{d}\eta^{\dagger}+\mathbf{d}^{*}\eta$, can strongly couple to the induced electric field of the microwave photon through electric-dipole coupling \cite{koch2007}
\begin{eqnarray}
H_{\text{ph-q}}=-\mathbf{E}_{\text{ph}}\cdot\hat{\mathbf{d}},
\end{eqnarray}
where $\mathbf{E}_{\text{ph}}=\frac{d\mathbf{A}_{\mathbf{k}}(\mathbf{r}, t)}{dt}$ determines the photon-induced electric field. If we assume $\mathbf{d}||\mathbf{e}_{R}$, then, under the rotating wave approximation, the photon-qubit interaction is described by the Hamiltonian 
\begin{eqnarray}
H_{\text{ph-q}}^{\mathbf{k}}=-g_{\text{ph-q}}^{\mathbf{k}}\eta^{\dagger}c_{\mathbf{k}}+{\rm H.c.},
\label{PhQH}
\end{eqnarray}
where the photon-qubit exchange coupling is given by 
\begin{eqnarray}
g_{\text{ph-q}}^{\mathbf{k}}=-id\omega_{c_{\mathbf{k}}}\exp[-i\mathbf{k} \cdot \mathbf{r}]
\end{eqnarray}
 with $d=|\mathbf{d}|$ being the strength of electric dipole of the superconducting transmon qubit.

Having specified each term in the Hamiltonian of Eq.~\eqref{MH}, we conclude that the magnon-photon-transmon hybrid system is explicitly described by the bosonized Hamiltonian    
\begin{eqnarray}
H_{\mathbf{k}}&=&\omega_{\alpha_{\mathbf{k}}}\alpha_{\mathbf{k}}^{\dagger}\alpha_{\mathbf{k}} +
\omega_{\beta_{-\mathbf{k}}}\beta_{-\mathbf{k}}^{\dagger} \beta_{-\mathbf{k}}
+\omega_{c_{\mathbf{k}}}c^{\dagger}_{\mathbf{k}}c_{\mathbf{k}}
+\omega_{q}\eta^{\dagger}\eta
\nonumber\\
&&
-\left[g_{\text{m-ph}}^{\mathbf{k}}c^{\dagger}_{\mathbf{k}}\alpha_{\mathbf{k}}
+g_{\text{ph-q}}^{\mathbf{k}}\eta^{\dagger}c_{\mathbf{k}}+ {\rm H.c.} \right],
\label{PhQH}
\end{eqnarray}
for a momentum $\mathbf{k}$ vector in the $z$-direction, the in-plane parallel photon polarization vector $\mathbf{e}_{R}$, and the superconducting dipole $\mathbf{d}||\mathbf{e}_{R}$.  

It is important to note that only the hybridized magnon in the $\alpha$ mode interacts with the photon and transmon modes in the Hamiltonian in 
Eq.~\eqref{PhQH}. In other words, the $\beta$ magnon mode is effectively decoupled from the other modes in the system. This is due to the fact that we use the right circularly polarized microwave cavity electromagnetic field, which only couples to the magnon with the same polarization, the $\alpha$ mode. On the one hand, if we use a left circularly polarized cavity field, it couples the $\beta$ magnon mode with the photon and the transmon modes, and instead leaves the $\alpha$ magnon mode decoupled from the rest of the system.

The hybrid quantum system described by Eq.~\eqref{PhQH} provides a promising platform to observe and verify quantum effects in quantum magnonics and exploit them for new quantum applications. Below we employ this hybrid platform to propose a new experimental setup for observing polarized twin magnon modes as well as intrinsic two-mode magnon entanglement in bipartite AFM materials via a transmon qubit. In the next section we briefly describe the basic concepts of two-mode entanglement in AFMs. 

\section{magnon-magnon entanglement} 
\label{magnon-magnon-entanglement}

 Let us focus on the two-mode magnon Hamiltonian described by $H_{\text{m}}^{\mathbf{k}}$ above. The coupling parameter $g_{\text{m-m}}^{\mathbf{k}}$ in Eq.~\eqref{MMH}, which is mainly given by the AFM coupling between the two opposite sublattices $A$ and $B$, introduce a strong squeezing and entanglement between bosonic magnon modes
 %, $a$ and $b$, 
 in a way that all the eigenstates of $H_{\text{m}}^{\mathbf{k}}$ become entangled in the Kittel $(a, b)$ modes  \cite{azimi-mousolou2020, azimi-mousolou2021}. Explicitly, the complete energy eigenbasis of the  
Hamiltonian $H_{\text{m}}^{\mathbf{k}}$ can be expressed in the following form  
\begin{eqnarray}
\ket{\psi_{xy}(r_{\mathbf{k}}, \phi_{\mathbf{k}})}&=&(\alpha^{\dagger}_{\mathbf{k}})^{x}(\beta^{\dagger}_{-\mathbf{k}})^{y}\ket{\psi_{00}(r_{\mathbf{k}}, \phi_{\mathbf{k}})}
\label{eq:MES}
\end{eqnarray}
for positive integers $x$ and $y$, and the two-mode squeezed vacuum ground state  
\begin{eqnarray}
\ket{\psi_{00}(r_{\mathbf{k}}, \phi_{\mathbf{k}})}= \frac{1}{\cosh r_{\mathbf{k}}} 
\sum_{n=0}^{\infty} e^{in\phi_{\mathbf{k}}} \tanh^{n} r_{\mathbf{k}} 
\ket{n; a_{\mathbf{k}}}\ket{n; b_{-\mathbf{k}}}
\nonumber\\
\label{eq: two-mode squeezing ground state}
\end{eqnarray}
given in the Kittel $(a, b)$ magnon basis. Here, $x$ and $y$ represent the number of magnons in the hybridized magnon modes $\alpha_{\mathbf{k}}$ and $\beta_{-\mathbf{k}}$, respectively. Note that the hybridized magnon modes $(\alpha, \beta)$ are related to the Kittel magnon modes $(a, b)$ through Eq.~\eqref{eq:FBT}.

\begin{figure}[h]
\begin{center}
\includegraphics[width=80mm, height=42mm]{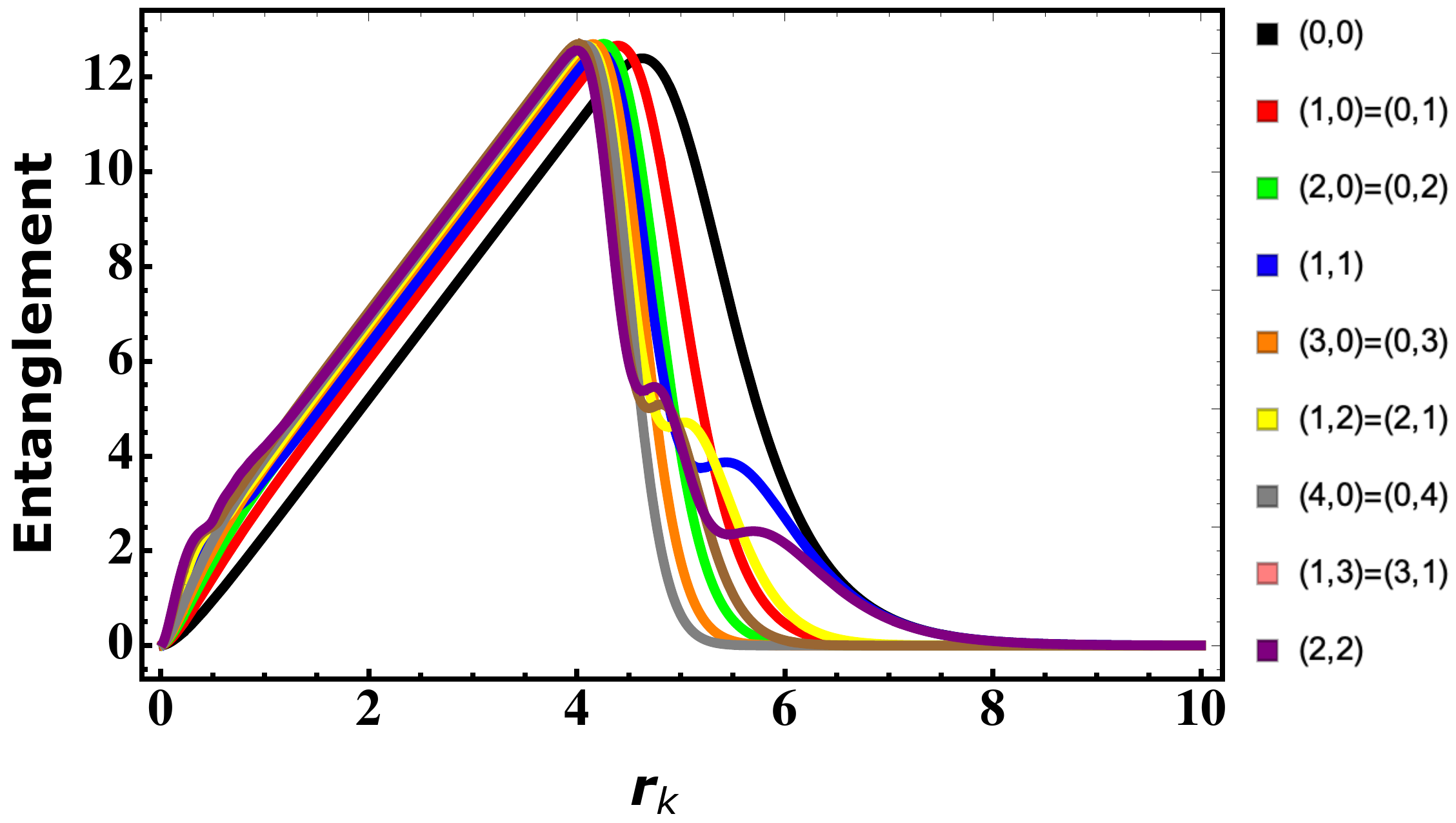}
\end{center}
\caption{(Color online) Entanglement of magnon eigenstates corresponding to pairs of magnon numbers $(x, y)$ against the entanglement (squeezing) parameter $r_{\mathbf{k}}$.}
\label{fig:Entanglement}
\end{figure}

Fig.~\ref{fig:Entanglement} 
illustrates the entropies of entanglement of the energy eigenbasis in Eq.~\eqref{eq:MES} for selected pairs of magnon numbers $(x, y)$ as functions of the squeezing parameter $r_{\mathbf{k}}$.
The squeezing parameter $r_{\mathbf{k}}$, which is given in Eq.\ \eqref{r-phi} by the ratio of the magnon-magnon coupling  $g_{\text{m-m}}^{\mathbf{k}}$ to the average single magnon energies in the Kittel modes, is actually the only parameter that determines the entropies of entanglement of the complete energy eigenbasis. This follows from the fact that the states in Eq.~\eqref{eq:MES} are determined by $(r_{\mathbf{k}}, \phi_{\mathbf{k}})$ and $\phi_{\mathbf{k}}$ contributes only to the phase factors of the Schmidt coefficients in the Schmidt decompositions of these states.

We remind the reader that the entropy of entanglement for a bipartite state $\ket{\psi}\in H_A\otimes H_B$ is given by 
\begin{eqnarray}
E\left[\ket{\psi}\right]&=&-\sum_{n}|\chi_{n}|^{2}\log|\chi_{n}|^{2}
\label{EEE}
\end{eqnarray}
with $\chi_{n}$'s being the Schmidt coefficients in $\ket{\psi}=\sum_{n}\chi_{n}\ket{i_n; A}\ket{j_n; B}$, where  $\ket{i_n; A}$ and $\ket{j_n; B}$ are orthonormal states in subsystem 
$A$ and subsystem $B$, respectively \cite{nielsen2000}.

For the energy eigenstates in 
Eq.~\eqref{eq:MES}, we obtain the following normalized Schmidt decompositions
\begin{eqnarray}
\begin{array}{ll}
\ket{\psi_{xy}(r_{\mathbf{k}}, \phi_{\mathbf{k}})} = &  \\ 
& \\
\left\{
\begin{array}{ll}
\sum_{n=0}^{\infty} p^{(x, y)}_{n; \mathbf{k}}\ket{n+\delta; a_{\mathbf{k}}} 
\ket{n; b_{-\mathbf{k}}},\  x\ge y \\
\sum_{n=0}^{\infty} p^{(x, y)}_{n; \mathbf{k}}\ket{n; a_{\mathbf{k}}} 
\ket{n+\delta; b_{-\mathbf{k}}},\ x\le y 
\end{array}
\right. 
&
\end{array}
\label{EES}
\end{eqnarray}
where $\delta=|x-y|$. Here, the Schmidt coefficients are given by
\begin{eqnarray}
p^{(x, y)}_{n; \mathbf{k}}=\frac{1}{\sqrt{x!y!}}
\left(\frac{1}{u_{\mathbf{k}}^{*}}\right)^{\delta}\left(\frac{1}{u_{\mathbf{k}}^{*}v_{\mathbf{k}}}\right)^{m}f^{(m, \delta)}_{n; \mathbf{k}}p_{n; \mathbf{k}},\ \ \ 
\label{SCES1}
\end{eqnarray}
for $m=\min\{x, y\}$, with 
\begin{eqnarray}
p_{n; \mathbf{k}} = \frac{e^{in\phi_{\mathbf{k}}}}{\cosh 
r_{\mathbf{k}}}\tanh^{n}r_{\mathbf{k}},
\label{SCES2}
\end{eqnarray}
and $f^{(m, \delta)}_{n; \mathbf{k}}$ that satisfies the following recursive relations 
\begin{eqnarray}
f^{(m, \delta>0)}_{n; \mathbf{k}}&=&|u_{\mathbf{k}}|^{2}\sqrt{n+\delta}f^{(m, \delta-1)}_{n; \mathbf{k}}
\nonumber \\  & & -|v_{\mathbf{k}}|^{2}\sqrt{n+1}f^{(m, \delta-1)}_{n+1; \mathbf{k}} , \nonumber\\
f^{(m>0, 0)}_{n; \mathbf{k}}&=&n|u_{\mathbf{k}}|^{4}f^{(m-1, 0)}_{n-1; \mathbf{k}}-(2n+1)|u_{\mathbf{k}}v_{\mathbf{k}}|^{2}f^{(m-1,0)}_{n; \mathbf{k}}\nonumber\\
&&+(n+1)|v_{\mathbf{k}}|^{4}f^{(m-1,0)}_{n+1; \mathbf{k}} 
\nonumber\\
\label{SCES3}
\end{eqnarray}
with $f^{(0, 0)}_{n; \mathbf{k}}=1$ for each $n$.  From Eqs. \eqref{SCES1}-\eqref{SCES3}, it is clear that the absolute value of the Schmidt coefficients $|p^{(x, y)}_{n; \mathbf{k}}|$,
and thus the entanglement entropies of all energy eigenbasis states in the Kittel magnon modes $(a, b)$, namely,
\begin{eqnarray}
E\left[(\alpha^{\dagger}_{\mathbf{k}})^{x}(\beta^{\dagger}_{-\mathbf{k}})^{y}\ket{\psi_{00}}\right]&=&-\sum_{n=0}^{\infty}|p^{(x, y)}_{n; \mathbf{k}}|^{2}\log|p^{(x, y)}_{n; \mathbf{k}}|^{2},\nonumber\\
\label{EEE}
\end{eqnarray}
are single variable functions of the squeezing parameter $r_{\mathbf{k}}$. In other words, the squeezing parameter $r_{\mathbf{k}}$ is the only entanglement parameter that determines two-mode magnon entanglement in the AFM system described by $H_{\text{m}}^{\mathbf{k}}$.

In the following we show how a superconducting transmon qubit can be used to observe different magnons and the squeezing/entanglement parameter $r_{\mathbf{k}}$. The latter allows us to quantify quantum characteristics such as two-mode squeezing and entanglement in AFM materials.

\section{Sensing magnons and thier quantum characteristics with transmons} 
\label{Sensing-magnons-quantum-characteristics-with-transmons}
\subsection{Magnon-transmon effective coupling}  
The Hamiltonian in Eq.~\eqref{PhQH}, that allows for magnon-photon-transmon hybrid states,
provides an effective photon mediated magnon-transmon coupling. To determine this effective coupling rate one may use the Schrieffer–Wolff unitary transformation \cite{schrieffer1966}, 
\begin{eqnarray}
H^{\prime}_{\mathbf{k}}=e^{W_{\mathbf{k}}}H_{\mathbf{k}}e^{-W_{\mathbf{k}}}
\label{Schrieffer-Wolff}
\end{eqnarray}
to effectively decouple the photon mode from magnon and transmon modes in the hybrid Hamiltonian up to first order. 

Consider the following decomposition of the hybrid Hamiltonian in 
Eq.~\eqref{PhQH}
\begin{eqnarray}
H_{\mathbf{k}}&=&H_{\mathbf{k}; 0}+V_{\mathbf{k}},\nonumber\\
H_{\mathbf{k}; 0}&=&\omega_{\alpha_{\mathbf{k}}}\alpha_{\mathbf{k}}^{\dagger}\alpha_{\mathbf{k}} 
+\omega_{c_{\mathbf{k}}}c^{\dagger}_{\mathbf{k}}c_{\mathbf{k}}
+\omega_{q}\eta^{\dagger}\eta ,
\nonumber\\
V_{\mathbf{k}}&=&-g_{\text{m-ph}}^{\mathbf{k}}c^{\dagger}_{\mathbf{k}}\alpha_{\mathbf{k}}
-g_{\text{ph-q}}^{\mathbf{k}}\eta^{\dagger}c_{\mathbf{k}}+ {\rm H.c.},
\label{EffPhQH}
\end{eqnarray}
where we neglect the magnon $\beta$ mode as it is decoupled from the rest of the Hamiltonian $H_{\mathbf{k}}$.
By using the Baker-Campbell-Haussdorf formula, 
the transformation in 
Eq.~\eqref{Schrieffer-Wolff} can be expanded as
\begin{eqnarray}
 H^{\prime}_{\mathbf{k}} & = & H_{\mathbf{k}; 0}+V_{\mathbf{k}}+[W_{\mathbf{k}}, H_{\mathbf{k}; 0}]+[W_{\mathbf{k}}, V_{\mathbf{k}}]
\nonumber\\
& & + \frac{1}{2}[W_{\mathbf{k}}, [W_{\mathbf{k}}, H_{\mathbf{k}; 0}]]+\frac{1}{2}[W_{\mathbf{k}}, [W_{\mathbf{k}}, V_{\mathbf{k}}]]+\ldots \ \ \ \     
\label{Baker-Campbell-Haussdorf}
\end{eqnarray}
 This three-mode Schrieffer–Wolff Hamiltonian can be made block diagonal turning the system into a two-mode magnon-transmon subsystem decoupled from a one-mode photon subsystem
 by choosing the generator $W_{\mathbf{k}}$ such that
 \begin{eqnarray}
V_{\mathbf{k}}+[W_{\mathbf{k}}, H_{\mathbf{k}; 0}]=0.
\label{generator}
 \end{eqnarray}
 By substituting the solution of Eq.\ \eqref{generator} into 
 Eq.~\eqref{Baker-Campbell-Haussdorf}, one can obtain the 
 standard form of the Schrieffer–Wolff Hamiltonian 
\begin{eqnarray}
%AD
H^{\prime}_{\mathbf{k}}=
H_{\mathbf{k}; 0}+
{\frac{1}{2}}[W_{\mathbf{k}}, V_{\mathbf{k}}]+
O(V_{\mathbf{k}}^{3})
\end{eqnarray}
 up to first order in the interaction term $V_{\mathbf{k}}$.

Equation \eqref{generator} always has a definite solution as the perturbative component $V_{\mathbf{k}}$ is off-diagonal in the eigenbasis of 
$H_{\mathbf{k}; 0}$. By solving Eq.\ \eqref{generator}, we obtain the generator of the Schrieffer–Wolff transformation
 \begin{eqnarray}
W_{\mathbf{k}}=\left[\frac{g_{\text{m-ph}}^{\mathbf{k}}}{\omega_{\alpha_{\mathbf{k}}}-\omega_{c_{\mathbf{k}}}}c_{\mathbf{k}}^{\dagger}\alpha_{\mathbf{k}}
-\frac{g_{\text{ph-q}}^{\mathbf{k}}}{\omega_{q}-\omega_{c_{\mathbf{k}}}}\eta^{\dagger}c_{\mathbf{k}}\right]- {\rm H.c.}\ \ \ 
 \end{eqnarray}
that leads to the following block diagonal hybrid Hamiltonian
 \begin{eqnarray}
H^{\prime}_{\mathbf{k}}&=&H_{\mathbf{k}; 0}+{\frac {1}{2}}[W_{\mathbf{k}}, V_{\mathbf{k}}]
\nonumber\\
&=&\omega^{\prime}_{c_{\mathbf{k}}}c^{\dagger}_{\mathbf{k}}c_{\mathbf{k}}+\omega^{\prime}_{\alpha_{\mathbf{k}}}\alpha_{\mathbf{k}}^{\dagger}\alpha_{\mathbf{k}} 
+\omega^{\prime}_{q}\eta^{\dagger}\eta
\nonumber\\
&&+g_{\text{m-q}}^{\mathbf{k}}\eta^{\dagger}_{\mathbf{k}}\alpha_{\mathbf{k}}
+\left(g_{\text{m-q}}^{\mathbf{k}}\right)^{*}\alpha_{\mathbf{k}}^{\dagger}\eta_{\mathbf{k}},
\label{SWPhQH}
 \end{eqnarray}
where
\begin{eqnarray}
\omega^{\prime}_{c_{\mathbf{k}}}&=&\omega_{c_{\mathbf{k}}}-\frac{|g_{\text{m-ph}}^{\mathbf{k}}|^{2}}{\omega_{\alpha_{\mathbf{k}}}-\omega_{c_{\mathbf{k}}}}-\frac{|g_{\text{ph-q}}^{\mathbf{k}}|^{2}}{\omega_{q}-\omega_{c_{\mathbf{k}}}} ,
\nonumber\\
\omega^{\prime}_{\alpha_{\mathbf{k}}}&=&\omega_{\alpha_{\mathbf{k}}}+\frac{|g_{\text{m-ph}}^{\mathbf{k}}|^{2}}{\omega_{\alpha_{\mathbf{k}}}-\omega_{c_{\mathbf{k}}}} ,
\nonumber\\
\omega^{\prime}_{q}&=&\omega_{q}+\frac{|g_{\text{ph-q}}^{\mathbf{k}}|^{2}}{\omega_{q}-\omega_{c_{\mathbf{k}}}} ,
\nonumber\\
g_{\text{m-q}}^{\mathbf{k}}&=&g_{\text{m-ph}}^{\mathbf{k}}g_{\text{ph-q}}^{\mathbf{k}}\left[\frac{1}{\omega_{\alpha_{\mathbf{k}}}-\omega_{c_{\mathbf{k}}}}+\frac{1}{\omega_{q}-\omega_{c_{\mathbf{k}}}}\right].
 \end{eqnarray}
As the photon mode is effectively decoupled from the rest of the Hamiltonian in Eq.~\eqref{SWPhQH}, the effective magnon-transmon interacting Hamiltonian reads 
\begin{eqnarray}
H^{\mathbf{k}; \text{eff}}_{m-q}&=&\omega^{\prime}_{\alpha_{\mathbf{k}}}\alpha_{\mathbf{k}}^{\dagger}\alpha_{\mathbf{k}} 
+\omega^{\prime}_{q}\eta^{\dagger}\eta
\nonumber \\ 
& & +g_{\text{m-q}}^{\mathbf{k}}\eta^{\dagger}_{\mathbf{k}}\alpha_{\mathbf{k}}
+\left(g_{\text{m-q}}^{\mathbf{k}}\right)^{*}\alpha_{\mathbf{k}}^{\dagger}\eta_{\mathbf{k}}.
\label{eq:eff-m-q-Hamiltonian}
 \end{eqnarray}

\subsection{Transmon-qubit to probe magnons and their quantum characteristics in AFMs}  

The computational basis of the transmon qubit consits of the ground and first excited states  $\ket{0}\equiv \ket{g}$ and $\ket{1}\equiv \ket{e}$, respectively, of the anharmonic oscillator in the transmon regime. 
In this case, the raising and lowering operators of the transmon qubit can be represented as $\eta^{\dagger}=\ket{1}\bra{0}$ and $\eta=\ket{0}\bra{1}$. The eigenstates of the  number operator 
\begin{eqnarray}
N_{\mathbf{k}}=\alpha_{\mathbf{k}}^{\dagger}\alpha_{\mathbf{k}} 
+\eta^{\dagger}\eta=\alpha_{\mathbf{k}}^{\dagger}\alpha_{\mathbf{k}} +\ket{1}\bra{1},
 \end{eqnarray}
are $\{\ket{0, 0}; \ket{1, 0}, \ket{0, 1};...; \ket{n, 0}, \ket{n-1, 1};...\}$, where the first entry counts the number of magnons in the hybridized mode $\alpha$ and the second entry labels the qubit state. These eigenstates span the magnon-qubit Hilbert space. The number operator commutes with the effective Hamiltonian in Eq.~\eqref{eq:eff-m-q-Hamiltonian}, i.e., 
\begin{eqnarray}
[N_{\mathbf{k}}, H^{\mathbf{k}; \text{eff}}_{m-q}]=0. 
 \end{eqnarray} 
This implies that the effective Hamiltonian takes the block diagonal form:
\begin{eqnarray}
H^{\mathbf{k}; \text{eff}}_{m-q}=\bigoplus_{n=0}H^{\mathbf{k}; n}_{m-q},
 \end{eqnarray}
where $n$ is the eigenvalue of the number operator $N_{\mathbf{k}}$, i.e., counts the total number of magnon and transmon excitations.
Except for the case $n=0$ that the Hamiltonian submatrix is a 1D block, for each $n>0$ the block Hamiltonians $H^{\mathbf{k}; n}_{m-q}$ are $2\times 2$ matrix of the form 
 \begin{eqnarray}
H^{\mathbf{k}; n}_{m-q}=\left(
\begin{array}{cc}
n \omega^{\prime}_{\alpha_{\mathbf{k}}}& \sqrt{n}\left(g_{\text{m-q}}^{\mathbf{k}}\right)^{*}   \\
\sqrt{n}g_{\text{m-q}}^{\mathbf{k}} & n \omega^{\prime}_{\alpha_{\mathbf{k}}}-2\Delta_{\mathbf{k}}      
\end{array}
\right),
\label{eq:block Hamiltonian}
\end{eqnarray}
with $\Delta_{\mathbf{k}}=\left(\omega^{\prime}_{\alpha_{\mathbf{k}}}-\omega^{\prime}_{q}\right)/2$ being the detuning between magnon and qubit frequencies. 

By shifting the qubit energy levels $\ket{0}$ and $\ket{1}$ with the amount of $\Delta_{\mathbf{k}}$, we may rewrite the Hamiltonian in 
Eq.~\eqref{eq:block Hamiltonian} as a effective single transmon qubit Hamiltonian
 \begin{eqnarray}
H^{\text{eff}}_{q}=n \omega^{\prime}_{\alpha_{\mathbf{k}}}\mathbb{I}+\sqrt{n}\Omega_{\mathbf{k}}^x\sigma_x
+\sqrt{n}\Omega_{\mathbf{k}}^y\sigma_y+\Delta_{\mathbf{k}}\sigma_z
\label{eq:effectiveQH}
\end{eqnarray}
for each $n$. Here, $\Omega_{\mathbf{k}}=\Omega_{\mathbf{k}}^x+i\Omega_{\mathbf{k}}^y=g_{\text{m-q}}^{\mathbf{k}}$ characterizes the Rabi frequency of the qubit, $\mathbb{I}$ is the $2\times 2$ identity matrix and $\sigma_l,\ \ l=x, y, z$, are the Pauli matrices in the ordered effective qubit basis $\{\ket{n, 0}, \ket{n-1, 1}\}$.
This Hamiltonian results in the following energy eigensystem:
 \begin{eqnarray}
\epsilon_{\pm}&=&n \omega^{\prime}_{\alpha_{\mathbf{k}}}\pm\sqrt{\Delta_{\mathbf{k}}^2+n|\Omega_{\mathbf{k}}|^2},
\nonumber\\
\ket{\epsilon_{+}}&=&\cos\left(\frac{\theta_{\mathbf{k}}}{2}\right)\ket{n, 0}+ e^{i\phi_{\mathbf{k}}}\sin\left(\frac{\theta_{\mathbf{k}}}{2}\right)\ket{n-1, 1},
\nonumber\\
\ket{\epsilon_{-}}&=&\sin\left(\frac{\theta_{\mathbf{k}}}{2}\right)\ket{n, 0}- e^{i\phi_{\mathbf{k}}}\cos\left(\frac{\theta_{\mathbf{k}}}{2}\right)\ket{n-1, 1}\ \ \ \ 
\end{eqnarray}
with $\Omega_{\mathbf{k}}=|\Omega_{\mathbf{k}}|e^{i\phi_{\mathbf{k}}}$ and $\tan\theta_{\mathbf{k}}=\frac{|\Omega_{\mathbf{k}}|}{\Delta_{\mathbf{k}}}$.

Suppose the transmon qubit is initialized in the state $\ket{0}$ at time $t=0$ for a fixed $n$, for instance $n=1$, that is 
$\ket{\psi(0)}=\ket{1,0}$. Governed by the effective qubit Hamiltonian in 
Eq.~\eqref{eq:effectiveQH},
the initial state evolves to
 \begin{eqnarray}
\ket{\psi(t)}&=&e^{-itH^{\text{eff}}_{q}}\ket{\psi(0)}\nonumber\\
&=&e^{-it\epsilon_{+}}\cos\left(\frac{\theta_{\mathbf{k}}}{2}\right)\ket{\epsilon_{+}}+e^{-it\epsilon_{-}}\sin\left(\frac{\theta_{\mathbf{k}}}{2}\right)\ket{\epsilon_{-}},\nonumber\\
\end{eqnarray}
after time $t$, which give rise to the following Rabi oscillation 
 \begin{eqnarray}
P_{0\rightarrow 1}(t)&=&|\langle1|\psi(t)\rangle|^{2}=\sin^{2}\left(\theta_{\mathbf{k}}\right)\sin^{2}\left(\frac{(\epsilon_{+}-\epsilon_{-})t}{2}\right)\nonumber\\
&=&\frac{|\Omega_{\mathbf{k}}|^2}{\Delta_{\mathbf{k}}^2+|\Omega_{\mathbf{k}}|^2}\sin^{2}\left(\frac{(\epsilon_{+}-\epsilon_{-})t}{2}\right).
\end{eqnarray}
This indicates that the probability of finding the transmon qubit
in the state $\ket{1}$ after time $t$ oscillates with the frequency  
 \begin{eqnarray}
f_{\mathbf{k}}=\frac{(\epsilon_{+}-\epsilon_{-})}{2}=\sqrt{\Delta_{\mathbf{k}}^2+|\Omega_{\mathbf{k}}|^2},
\end{eqnarray}
and intensity 
\begin{eqnarray}
I_{\mathbf{k}}=\frac{|\Omega_{\mathbf{k}}|^2}{\Delta_{\mathbf{k}}^2+|\Omega_{\mathbf{k}}|^2}.
\end{eqnarray}
Note that the maximum intensity $I_{\mathbf{k}}=1$ occurs at the zero detuning $\Delta_{\mathbf{k}}=0$, which is equivalent to the following qubit parameter tuning 
\begin{eqnarray}
\omega_q&=&\omega_{\alpha_{\mathbf{k}}},\nonumber\\
|g_{\text{ph-q}}^{\mathbf{k}}|&=&|g_{\text{m-ph}}^{\mathbf{k}}|.
\label{QP-zerodetuning}
\end{eqnarray}
The detuning can be achieved, for instance, by appropriate adjustments of photon frequency and amplitude of vector potential as well as an applied magnetic 
field in the $z$ direction, as depicted in Fig.\ \ref{fig:model}.
As a result of zero detuning, the angular frequency of the Rabi oscillation becomes 
 \begin{eqnarray}
f_{\mathbf{k}}=\frac{2|g_{\text{m-ph}}^{\mathbf{k}}|^2}{|\omega_{q}-\omega_{c_{\mathbf{k}}}|}=\frac{2\lambda_{\mathbf{k}}^2}{|\omega_{q}-\omega_{c_{\mathbf{k}}}|}\Delta\left[\psi_{00}(r_{\mathbf{k}}, \phi_{\mathbf{k}})\right],
\label{AFofRO}
\end{eqnarray}
where 
 \begin{eqnarray}
 \Delta\left[\psi_{00}(r_{\mathbf{k}}, \phi_{\mathbf{k}})\right]=\cosh 2r_{\mathbf{k}}+\sinh2r_{\mathbf{k}}\cos\phi_{\mathbf{k}}
 \end{eqnarray}
 is the the Einstein-Podolsky-Rosen (EPR) function for the two-mode ground state 
$\ket{\psi_{00}(r_{\mathbf{k}}, \phi_{\mathbf{k}})}$
given by Eq.~\eqref{eq: two-mode squeezing ground state} \cite{azimi-mousolou2020,azimi-mousolou2021} (see appendix for details about EPR). 
The EPR function, which characterizes the Bell-type nonlocal correlations known as EPR nonlocality, is a highly relevant concept in the study of continuous variable entanglement \cite{giedke2003, fadel2020}. 

\begin{figure}[h]
\begin{center}
\includegraphics[width=80mm, height=50mm]{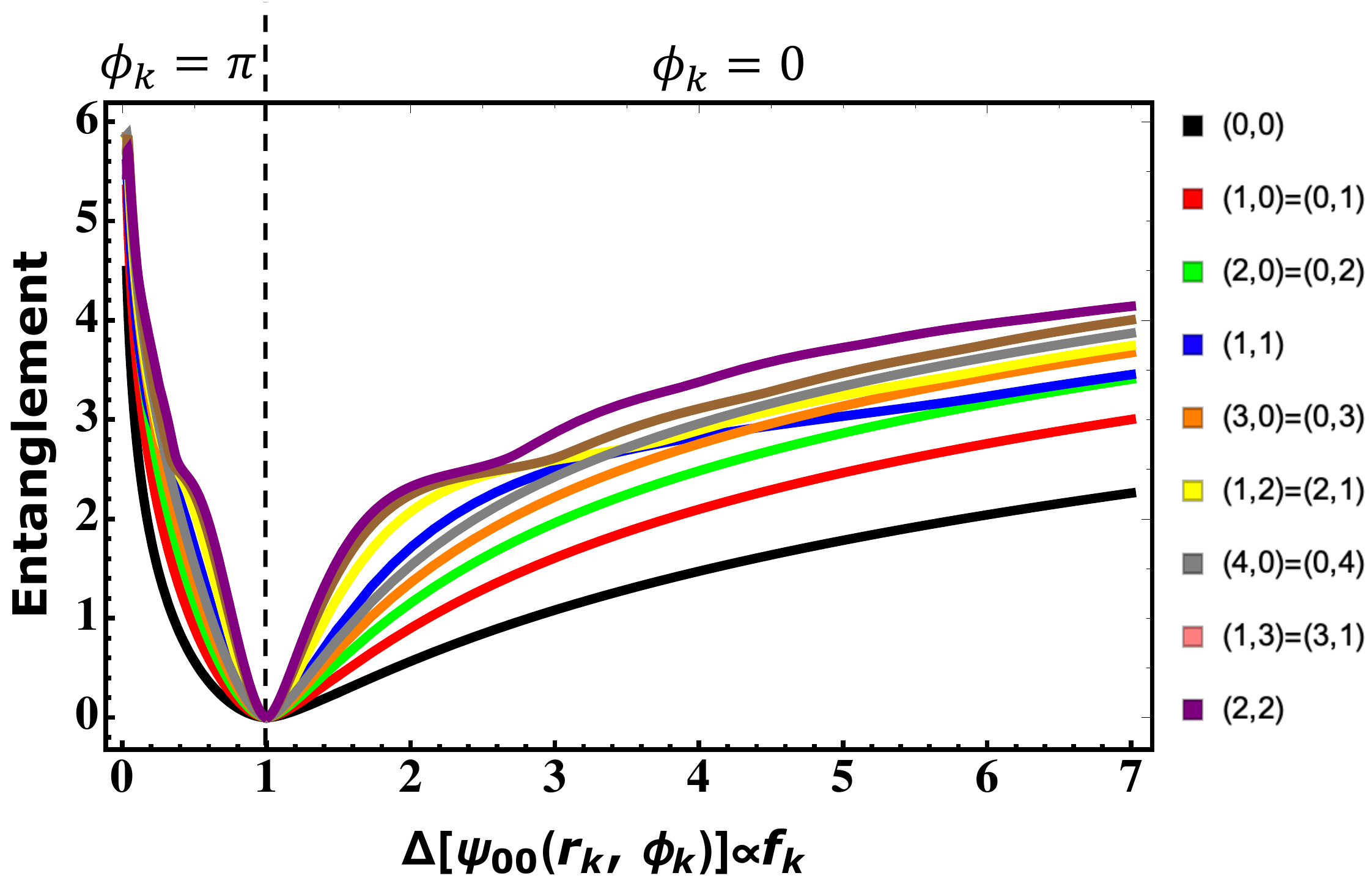}
\end{center} 
\caption{(Color online) Entanglement entropies of magnon eigenstates
corresponding to selected pairs of magnon numbers (x, y) against the EPR function $ \Delta\left[\psi_{00}(r_{\mathbf{k}}, \phi_{\mathbf{k}})\right]$ for AFM spin lattices. Stronger entanglement is observed for non-local states associated to  $\phi_{\mathbf{k}}=\pi$, whereas $\phi_{\mathbf{k}}=0$ represents a local state regime with weaker magnon-magnon entanglement.}
\label{fig:EvsEPR}
\end{figure}

We can always assume the parameter $\Gamma_{\mathbf{k}}$ in Eq.~\eqref{r-phi}
to be real-valued, in which case $\phi_{\mathbf{k}} = 0\ \text{or}\ \pi$ and thus
 \begin{eqnarray}
 \Delta\left[\psi_{00}(r_{\mathbf{k}}, \phi_{\mathbf{k}})\right] = 
\left\{ \begin{array}{ll} 
e^{2r_{\mathbf{k}}}, &\ \  \mathrm{if} \ \ \ \ \phi_{\mathbf{k}} = 0\ (\Gamma_{\mathbf{k}} < 0), \\ 
e^{-2r_{\mathbf{k}}}, &\ \  \mathrm{if} \ \ \ \ \phi_{\mathbf{k}} =\pi\ (\Gamma_{\mathbf{k}} > 0).
\end{array} \right. \nonumber\\
\label{EPRforEE}
\end{eqnarray}
Since the ground state EPR function and the magnon-magnon entanglement entropies all depend on the same entanglement (squeezing) parameter, one may observe the magnon-magnon entanglement through the EPR function 
$\Delta\left[\psi_{00}(r_{\mathbf{k}}, \phi_{\mathbf{k}})\right]$
 and in fact through the qubit angular frequency in Eq.~\eqref{AFofRO} of the Rabi oscillation. For instance, we obtain the entanglement entropy for the two-mode ground state
\begin{eqnarray}
E\left[\ket{\psi_{00}(r_{\mathbf{k}}, \phi_{\mathbf{k}})}\right] & = & \left[ \cosh^{2} (r_{\mathbf{k}}) \log \cosh^{2} (r_{\mathbf{k}}) \right. 
\nonumber\\
& & \left. -\sinh^{2}(r_{\mathbf{k}})\log\sinh^{2}(r_{\mathbf{k}})\right],
\label{entropymeasure}
\end{eqnarray}
as a function of the qubit angular frequency through 
\begin{eqnarray}
r_{\mathbf{k}} & = & \frac{e^{i\phi_{\mathbf{k}}}}{2}\Delta\left[\psi_{00}(r_{\mathbf{k}}, \phi_{\mathbf{k}})\right]=
\frac{e^{i\phi_{\mathbf{k}}}|\omega_{q}-\omega_{c_{\mathbf{k}}}|}{4\lambda_{\mathbf{k}}^2}f_{\mathbf{k}}
\label{SPforEE}
\end{eqnarray}
for $\phi_{\mathbf{k}}=0,\ \pi$.  
Eq. \eqref{SPforEE} follows from Eqs. \eqref{AFofRO} and \eqref{EPRforEE}. The entanglement entropies of all magnon eigenbasis states given by Eq. \eqref{EEE} are actually functions of the qubit angular frequency through the relation in Eq. \eqref{SPforEE}.
In practice the entanglement entropy, Eq.\ \eqref{entropymeasure}, is a function of the parameter $r_{\mathbf{k}}$, which can be identified by Eq.\ \eqref{SPforEE} once the qubit angular frequency $f_{\mathbf{k}}$ has been determined experimentally.
Figure~\ref{fig:EvsEPR} illustrates, as an example, the two-mode magnon entanglement in the ground (vacuum) state and number of excited states against the EPR function $ \Delta\left[\psi_{00}(r_{\mathbf{k}}, \phi_{\mathbf{k}})\right]\propto f_{\mathbf{k}}$, for AFM spin lattices.

Two distinct regions, the non-local bipartite entangled state, $\phi_{\mathbf{k}}=\pi$, and the local bipartite entangled state, $\phi_{\mathbf{k}}=0$, with transition point at $ \Delta\left[\psi_{00}(r_{\mathbf{k}}, \phi_{\mathbf{k}})\right]=1$ can be distinguished in Fig.~\ref{fig:EvsEPR}. The region of stronger magnon-magnon entanglement for non-local two-mode magnon state is observed by the EPR uncertainty relation $ \Delta\left[\psi_{00}(r_{\mathbf{k}}, \phi_{\mathbf{k}})\right]<1$. 
The clear relation between the EPR function and the two-mode magnon entanglement entropy allows for experimental quantification of magnon-magnon entanglement through the EPR function $ \Delta\left[\psi_{00}(r_{\mathbf{k}}, \phi_{\mathbf{k}})\right]$ and indeed the  
frequency, $f_{\mathbf{k}}$, of Rabi oscillation of the transmon qubit. It is worth mentioning that the EPR nonlocality has been used for verification of entanglement in optical and atomic systems based on homodyne detection and types of interferometry setups \cite{gross2011, armstrong2015, peise2015, lee2016, kunkel2018, fadel2018, li2020}. However, these types of measurement setups are not realistic for magnon systems, since these technologies are mainly based on beam splitters that have limitations for characterizing magnon entanglement. We propose as a solution, a mechanism and measurement setup that rely on qubit-light-matter interaction as a probe to observe the EPR function and thus EPR nonlocality and the degree of magnon-magnon entanglement. Moreover, Eq.~\eqref{QP-zerodetuning} shows that at the zero 
detuning, the magnon frequency in the hybridized $\alpha$ mode can also be observed through qubit frequency. 

\begin{figure}[h]
\begin{center}
\includegraphics[width=86mm, height=33mm]{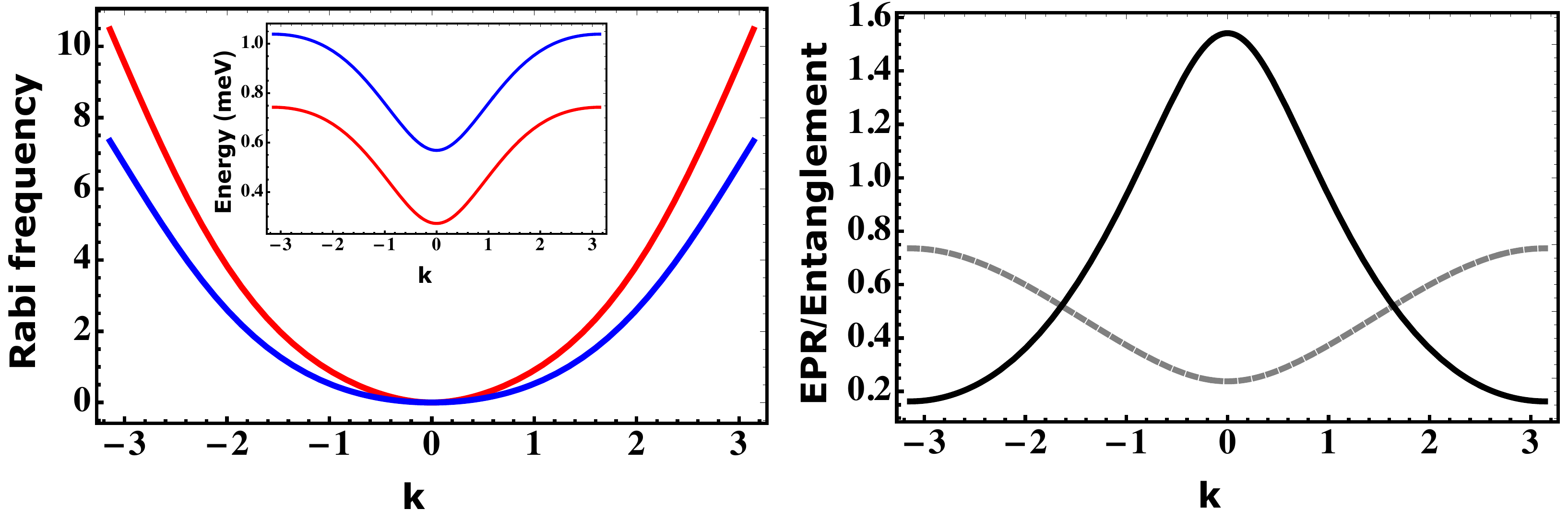}
\end{center} 
\caption{(Color online) Left panel: The angular frequency, $f_{\mathbf{k}}$, of the Rabi oscillation of transmon qubit depending on whether the transmon qubit is coupled to the magnon in $\alpha$ mode (red) through right circularly polarized photon or to the magnon in $\beta$ mode (blue) through left circularly polarized photon. The inset shows the corresponding dispersion energies for the two hybridized magnon modes $\alpha$ (red) and  $\beta$ (blue). Right panel: EPR function (gray dashed curve) and 
entanglement (black solid curve)  between Kittel magnon modes in the vacuum ground state for different values of lattice momentum $k$.
Similar results can be obtained for excited states. We assume 
 uniaxial AFM materials \cite{azimi-mousolou2021} with simple cubic lattice structure subjected to external magnetic field in the $z$ direction. The lattice momentum $k$ takes its values along $(0,0,1)$ direction with the lattice constant set to unity. We consider the nearest neighbor Heisenberg interaction $J$ and the easy-axis anisotropy $\mathcal{K}_z$ with model parameter values: $J=10 meV$, $\mathcal{K}_z=0.01J$, $B=2.5 T$ for the amplitude of the magnetic field in $z$ direction, and $S=1/2$. For the microwave cavity photon we assume $A_0=1 meV$ and $\omega_c=0.05 meV$.}
\label{fig:RFDofDM}
\end{figure}

A similar procedure and formulation as above hold if we couple the transmon qubit to a bipartite AFM material instead through the magnon $\beta$ mode, for instance, by using oppositely (left) circularly polarized light. Using different polarization for the photon would allow one to detect the twin chiral magnon modes in bipartite AFM materials.
Fig.\ \ref{fig:RFDofDM} shows that the angular frequency $f_{\mathbf{k}}$ of the Rabi oscillation of a transmon qubit can observe and distinguish the two hybridized magnong modes in the system provided that appropriate polarized light is used. The figure also shows the correlation between indistinguishablity of the two hybridized magnong modes, EPR nonlocality, and the entanglement between Kittel
magnon modes. The higher the indistinguishability (around the zone center), the higher the non-locality and entanglement. 

\section{conclusion}
\label{conclusion}

In conclusion, we demonstrate microwave cavity mediated hybridization
of superconducting transmon qubit and chiral magnons in bipartite AFM materials. We derive analytical expressions for the hybridized Hamiltonian and the coupling strengths. This coupling allows us not only to identify magnons in AFM materials, but also to verify their chirality and to characterize the nonlocality and bipartite entanglement between Kittel magnon modes in the system. These are all observed through measurement of the angular frequency of Rabi oscillation in the transmon qubit. We hope the present work opens up a new route to experimentally access rich quantum properties of magnons in AFM materials.   
The broad range of crystalline and synthetic AFM materials, such as the oxides NiO and MnO, the fluorides MnF$_2$ and FeF$_2$, 2D Ising systems like MnPSe$_3$, YIG-based synthetic AFMs, and perovskite manganites \cite{Jie2018,Takashi2016,Haakon2019,Thuc2021,Sheng2021, Changting2021, Rini2007, Ulbrich2011, rezende2019}, provide a space for experimental observation of the present results.

\section*{acknowledgments}
The authors acknowledge financial support from Knut and 
Alice Wallenberg Foundation through Grant No. 2018.0060. A.D. acknowledges financial support from the Swedish Research Council (VR) through Grants No.~2016-05980 and VR 2019-05304. O.E. acknowledges support from the Swedish Research Council (VR), the Swedish Foundation for Strategic Research (SSF), the Swedish Energy Agency (Energimyndigheten), ERC (synergy grant FASTCORR, project 854843), eSSENCE, and STandUPP. D.T. acknowledges support from the Swedish Research 
Council (VR) through Grant No. 2019-03666. E.S. acknowledges financial support from 
the Swedish Research Council (VR) through Grant No. 2017-03832. Some of the 
computations were performed on resources provided by the Swedish 
National Infrastructure for Computing (SNIC) at the National Supercomputer Center (NSC), 
Link\"oping University, the PDC Centre for High Performance Computing (PDC-HPC), KTH, 
and the High Performance Computing Center North (HPC2N), Ume{\aa} University.

\section*{Appendix} 
Here, for a general two-mode quantum state $\ket{\psi}$, the EPR function is quantified by \cite{giedke2003, fadel2020}
\begin{eqnarray}
\Delta(\psi) =   
\frac{1}{2}[\text{Var}_{\psi}(X_{\mathbf{k}}^{A}+X_{\mathbf{k}}^{B})
 + \text{Var}_{\psi}(P_{\mathbf{k}}^{A}-P_{\mathbf{k}}^{B})],
 \label{EPRR}
\end{eqnarray}
where $X_{\mathbf{k}}^{A} = \frac{a_{\mathbf{k}} + a_{\mathbf{k}}^{\dagger}}{\sqrt{2}}$ $\left(X_{\mathbf{k}}^{B} = 
\frac{b_{\mathbf{k}}+b_{\mathbf{k}}^{\dagger}}{\sqrt{2}}\right)$ and 
$P_{\mathbf{k}}^{A} = \frac{a_{\mathbf{k}} - a_{\mathbf{k}}^{\dagger}}{i\sqrt{2}}$ $\left(P_{\mathbf{k}}^{B} = \frac{b_{\mathbf{k}} - 
b_{\mathbf{k}}^{\dagger}}{i\sqrt{2}}\right)$ are assumed to be the dimensionless 
position and momentum quadratures for the $a_{\mathbf{k}} (b_{\mathbf{k}})$ mode, respectively. The 
$\text{Var}_{\psi}(V)$ is the variance of an Hermitian operator $V$ with respect to the state $\ket{\psi}$.
The uncertainty relation $\Delta(\psi)\ge 1$ is known to hold for any given bipartite separable state $\ket{\psi}$  \cite{fadel2020}. 
Therefore, any violation of this inequality is an indication of the state $\ket{\psi}$ being nonlocal and indeed a bipartite entangled state.
Note that the EPR nonlocality specifies a stronger type of entanglement than a nonzero entropy of entanglement in the sense that there are states with nonzero entropy of entanglement which do not violate the uncertainty relation.

\end{document}